\documentclass[aps,prl,twocolumn,showpacs,amsmath,preprintnumbers]{revtex4} 
\usepackage{amssymb,amsmath,graphicx}
\begin{document}
\title{Vacuum Fluctuations 
and the Small Scale Structure of Spacetime} 

\author{S.~Carlip}
\email[]{carlip@physics.ucdavis.edu}
\affiliation{Department of Physics, University of California, Davis, CA 95616, USA} 
\author{R.~A.\ Mosna}
\email[]{mosna@ime.unicamp.br} 
\author{J.~P.~M.~Pitelli}
\email[]{pitelli@ime.unicamp.br}
\affiliation{Instituto de Matem\'atica, Estat\'\i stica e 
        Computa\c{c}\~ao Cient\'\i fica, Universidade Estadual de Campinas,
        13083-859, Campinas, SP, Brazil}

\begin{abstract}
We show that vacuum fluctuations of the stress-energy tensor in 
two-dimensional dilaton gravity lead to a sharp focusing of light 
cones near the Planck scale, effectively breaking space up into a 
large number of causally disconnected regions.  This phenomenon, 
called ``asymptotic silence'' when it occurs in cosmology, might
help explain several puzzling features of quantum gravity, including
evidence of spontaneous dimensional reduction at short distances.
While our analysis focuses on a simplified two-dimensional model, 
we argue that the qualitative features should still be present 
in four dimensions.
\end{abstract}
\maketitle

\section {Introduction}

A fundamental goal of quantum gravity is to understand the causal
structure of spacetime---the behavior of light cones---at very small
scales.  One recent proposal \cite{Carlip1,Carlip2}, based on the strong
coupling approximation to the Wheeler-DeWitt equation \cite{Isham,%
Henneaux}, is that quantum effects may lead to strong focusing near 
the Planck scale, essentially collapsing light cones and breaking space 
up into a large number of very small, causally disconnected regions.  
Such a phenomenon occurs in classical cosmology near a spacelike 
singularity, where it has been called ``asymptotic silence'' 
\cite{Uggla}; it can be viewed as a kind of anti-Newtonian limit, 
in which the effective speed of light drops to zero.  Near the Planck 
scale, asymptotic silence could explain several puzzling features of 
quantum gravity, most notably the apparent dimensional reduction 
to two dimensions \cite{Carlip1,Carlip2,Stojkovic} that occurs in 
lattice models \cite{Ambjorn}, in some renormalization group analyses 
\cite{Reuter}, and elsewhere \cite{Modesto,Atick}

In cosmology, the strong focusing of null geodesics comes from the
presence of a singularity.  For quantum gravity, a different source is  
required.  Null geodesics are governed by the Raychaudhuri equation
\cite{Raychaudhuri,HawkingEllis}.  For a congruence of null geodesics%
---a pencil of light---with affinely parametrized null normals $\ell^a$, 
this equation tells us that
\begin{align}
\ell^a\nabla_a\theta 
    = -\frac{1}{2}\theta^2 - \sigma_{ab}\sigma^{ab}
    + \omega_{ab}\omega^{ab} - 8\pi T_{ab}\ell^a\ell^b ,
\label{a1}
\end{align}
where the expansion
\begin{align}
\theta = \frac{1}{A}\ell^a\nabla_a A
\label{a2}
\end{align} 
is the fractional rate of change of  area of a cross section of the 
pencil, $\sigma_{ab}$ is its shear, and $\omega_{ab}$ is its vorticity.
The stress-energy tensor $T_{ab}$ appears by virtue of the Einstein 
field equations, which relate it to the Ricci tensor.  (We 
use natural units, $\hbar = c = G_N = 1$.)   In particular, vacuum 
fluctuations with positive energy focus null geodesics, decreasing 
the expansion, while fluctuations with negative energy defocus them.

It has recently been shown that for many forms of matter, most vacuum
fluctuations have negative energy \cite{Fewster}.  These fluctuations 
have a strict lower bound, however, while the rarer positive fluctuations 
are unbounded.  A key question is then which of these dominate the
behavior of light cones near the Planck scale.

In this paper, we answer this question in the simplified context of 
two-dimensional dilaton gravity, a model obtained by dimensionally 
reducing general relativity to one space and one time dimension.  We 
show that the positive energy fluctuations win, and cause a collapse 
of light cones in a time on the order of 15 Planck times.  The
reduction to two dimensions is, for the moment, required for exact 
calculations; it is only in this setting that the spectrum of vacuum
fluctuations is fully understood.  But we show that the qualitative
features leading to strong focusing are also present in the full 
four-dimensional theory, strongly suggesting that our results should
apply to full general relativity.

\section{Dilaton gravity and the Raychaudhuri equation}

Our starting point is two-dimensional dilaton gravity, a theory that
can be described by the action \cite{Grumiller,%
Kunstatter}
\begin{align}
I =  \int d^2x\sqrt{|g|} \left[ \frac{1}{16\pi}\varphi R  
      + V[\varphi]  + \varphi L_m \right] ,
\label{b1}
\end{align}
where $\varphi$ is a scalar field, the dilaton, with a potential 
$V[\varphi]$ whose details will be unimportant.  This action can 
be obtained from standard general relativity by dimensional 
reduction, assuming either spherical or planar symmetry.  A 
conformal redefinition of fields is needed to bring the action into 
this form; the dilaton coupling to the matter Lagrangian $L_m$ 
is fixed provided the two-dimensional matter action is conformally 
invariant.  From now on, we assume such an invariance.

In two dimensions, the Raychaudhuri equation (\ref{a1}) does not
quite make sense, since there are no transverse directions in which 
to define the area or the expansion.  Its generalization to dilaton 
gravity is simple, however.  For any model of dilaton gravity obtained
by dimensional reduction, the dilaton has a direct physical 
interpretation as the transverse area in the ``missing'' dimensions.
We can therefore define a generalized expansion
\begin{align}
{\bar\theta} = \frac{1}{\varphi}\ell^a\nabla_a\varphi
     = \frac{d\ }{d\lambda}\ln\varphi ,
\label{b2}
\end{align}
where $\lambda$ is the affine parameter.  It is then an easy consequence 
of the dilaton gravity field equations that
\begin{align}
\frac{d\bar\theta}{d\lambda} 
     = -{\bar\theta}^2 - 8\pi T_{ab}\ell^a\ell^b
     = -{\bar\theta}^2 - 16\pi T_L ,
\label{b4}
\end{align}
where $T_L$ is the left-moving component of the stress-energy tensor.

We next need the vacuum fluctuations of the stress-energy tensor.  
For conformally invariant matter in two dimensions, Fewster, Ford, 
and Roman have found these exactly \cite{Fewster}.  A conformally
invariant field is characterized by a central charge $c$; a massless
scalar, for instance, has $c=1$.  To obtain a finite value for the 
stress-energy tensor, one must smear it over a small interval;
Fewster et al.\ use a Gaussian smearing function with width $\tau$.  
The probability distribution for the quantity $T_L$ in the Minkowski 
vacuum is then given by a shifted Gamma distribution 
\begin{align}
\mathop{Pr}(T_L = \omega) = 
     \vartheta(\omega + \omega_0)
     \frac{(\pi\tau^2)^\alpha(\omega + \omega_0)^{\alpha-1}}%
     {\Gamma(\alpha)} e^{-\pi\tau^2(\omega + \omega_0)} 
\label{b5}
\end{align}
where
\begin{align}
\omega_0 = \frac{c}{24\pi\tau^2}, \quad \alpha = \frac{c}{24}, 
\label{b6}
\end{align} 
and $\vartheta$ is the Heaviside step function.  As
noted earlier, this distribution is peaked at negative values of the
energy; for $c=1$, the probability of a positive fluctuation is only
$.16$.  There is, however, a long positive tail, as there must be in
order that the average $\langle T_L\rangle$ be zero.

Before proceeding with the calculation, it is useful to look at the
behavior of  the Raychaudhuri equation (\ref{b4}) with a constant
source $T_L = \omega$.  It is easy to see that
\begin{align}
{\bar\theta}(\lambda) = \left\{ \begin{array}{l}
       -\sqrt{16\pi\omega}\tan\sqrt{16\pi\omega}(\lambda-\lambda_0)\\[.7ex]
       \hspace*{3.75em} \hbox{if $\omega>0$} \\[1ex]
       \sqrt{|16\pi\omega|}\tanh\sqrt{|16\pi\omega|}(\lambda-\lambda_0)\\[.7ex]
       \hspace*{3.75em} \hbox{if $\omega<0$ and $|{\bar\theta}(0)|< \sqrt{|16\pi\omega|}$}\\[1ex]
       \pm\sqrt{|16\pi\omega|}\\[.7ex]
       \hspace*{3.75em} \hbox{if $\omega<0$ and ${\bar\theta}(0)= \pm\sqrt{|16\pi\omega|}$}\\[1ex]
       \sqrt{|16\pi\omega|}\coth\sqrt{|16\pi\omega|}(\lambda-\lambda_0)\\[.7ex]
       \hspace*{3.75em} \hbox{if $\omega\leq0$ and $|{\bar\theta}(0)|> \sqrt{|16\pi\omega|}$}
\end{array}\right.  
\label{b7}
\end{align}
where the integration constant $\lambda_0$ is determined from the initial
value of $\bar\theta$.  

As noted earlier, positive energy fluctuations can quickly drive the 
expansion to $-\infty$.  If $\bar\theta$ is initially negative, even small 
negative energy fluctuations---those on the $\coth$ branch of (\ref{b7})%
---cannot overcome the nonlinearities that also drive the expansion to
$-\infty$.  Larger negative energy fluctuations tend to defocus null 
geodesics, increasing the expansion, but even these have a limited effect.
The lower bound $-\omega_0$ for energy in the distribution (\ref{b5}) 
determines a maximum asymptotic value
\begin{align}
{\bar\theta}_+ = \sqrt{16\pi\omega_0} = \sqrt{2c/3\tau^2} ,
\label{b7b}
\end{align}
and if the expansion starts below this value, it will asymptote to at most
${\bar\theta}_+$.

Moreover, the same bound on negative energy fluctuations means that 
the positive contribution to the right-hand side of (\ref{b4}) cannot 
be larger than $16\pi\omega_0 = {\bar\theta}_+{}^2$.  The expansion 
thus has a critical negative value.  Any fluctuation that brings 
$\bar\theta$ to a value lower than
\begin{align}
{\bar\theta}_{\hbox{\scriptsize\it crit}} = -\sqrt{\frac{2c}{3\tau^2}}
     = -{\bar\theta}_+
\label{b8}
\end{align}
is irreversible: once the expansion becomes this negative, it will 
necessarily continue to decrease.  This gives a qualitative answer 
to our central question---in the long run, the positive energy 
fluctuations will always win, and the expansion will diverge to 
$-\infty$.  Note the crucial role of the nonlinear term in (\ref{b4});
a similar problem was considered in \cite{Borgman}, but only in
the approximation that $\bar\theta$ was small enough that this
term could be neglected.

To estimate the time to this ``collapse'' of the light cones, let us 
assume that the initial negative energy fluctuations push $\bar\theta$
to near its peak value of ${\bar\theta}_+$.  We can then ask for 
the probability of a positive energy fluctuation large enough to drive 
$\bar\theta$ to ${\bar\theta}_{\hbox{\scriptsize\it crit}}$ within 
a characteristic time $\tau$.  This can be determined from (\ref{b5}); 
for a massless scalar field ($c=1$), it is $\rho\approx .065$.  If we 
now treat these fluctuations as a Poisson process---i.e., independent 
events occurring randomly with probability $\rho\Delta t/\tau$ in any 
interval $\Delta t$---and ignore all smaller fluctuations, the time 
to ``collapse'' will be given by an exponential distribution 
$({\rho}/{\tau})e^{-\rho t/\tau}$, with a mean of approximately 
$15.4\tau$.   

This is, of course, an oversimplified picture of the combined effect of 
many vacuum fluctuations.  To obtain a more precise result, we next 
turn to a numerical analysis of the Raychaudhuri equation.

\section{Quantitative results}

Our analysis so far has been semiclassical: we consider quantum 
fluctuations of matter, but ignore purely quantum gravitational
effects.  We can, in principle, incorporate weak quantum fluctuations
of the metric into the stress-energy tensor in (\ref{b4}), but there
is no reason to trust our methods at scales at or below the Planck
scale.  We therefore choose the width $\tau$ of the Gaussian smearing
function to be the Planck length.   The overall system
(\ref{b4})--(\ref{b5}) is invariant under a simultaneous rescaling
of $\tau$ and the affine parameter $\lambda$, so this choice is not
critical; given any cutoff $\tau$, our results can be interpreted
as giving the focusing time in units of $\tau$.  For our matter
field, we choose a massless scalar, that is, a conformal field with
central charge $c=1$.

We proceed as follows:
\begin{enumerate} 
\item We choose an initial condition ${\bar\theta}_0 = 0$, and 
select a random value of the vacuum fluctuation $\omega$,
with a probability given by the distribution (\ref{b5}).  
\item We evolve forward one Planck time using (\ref{b7}), to determine
a new value ${\bar\theta}_1$.
\item Using ${\bar\theta}_1$ as an initial condition and choosing a
new value of $\omega$, we evolve another Planck time to determine
${\bar\theta}_2$.   We repeat the process, keeping track of the number 
of iterations, until we land on a negative branch of (\ref{b7}) with  the 
expansion diverging to $-\infty$ during the step.
\end{enumerate}

\begin{figure}
\includegraphics[width=3.4in]{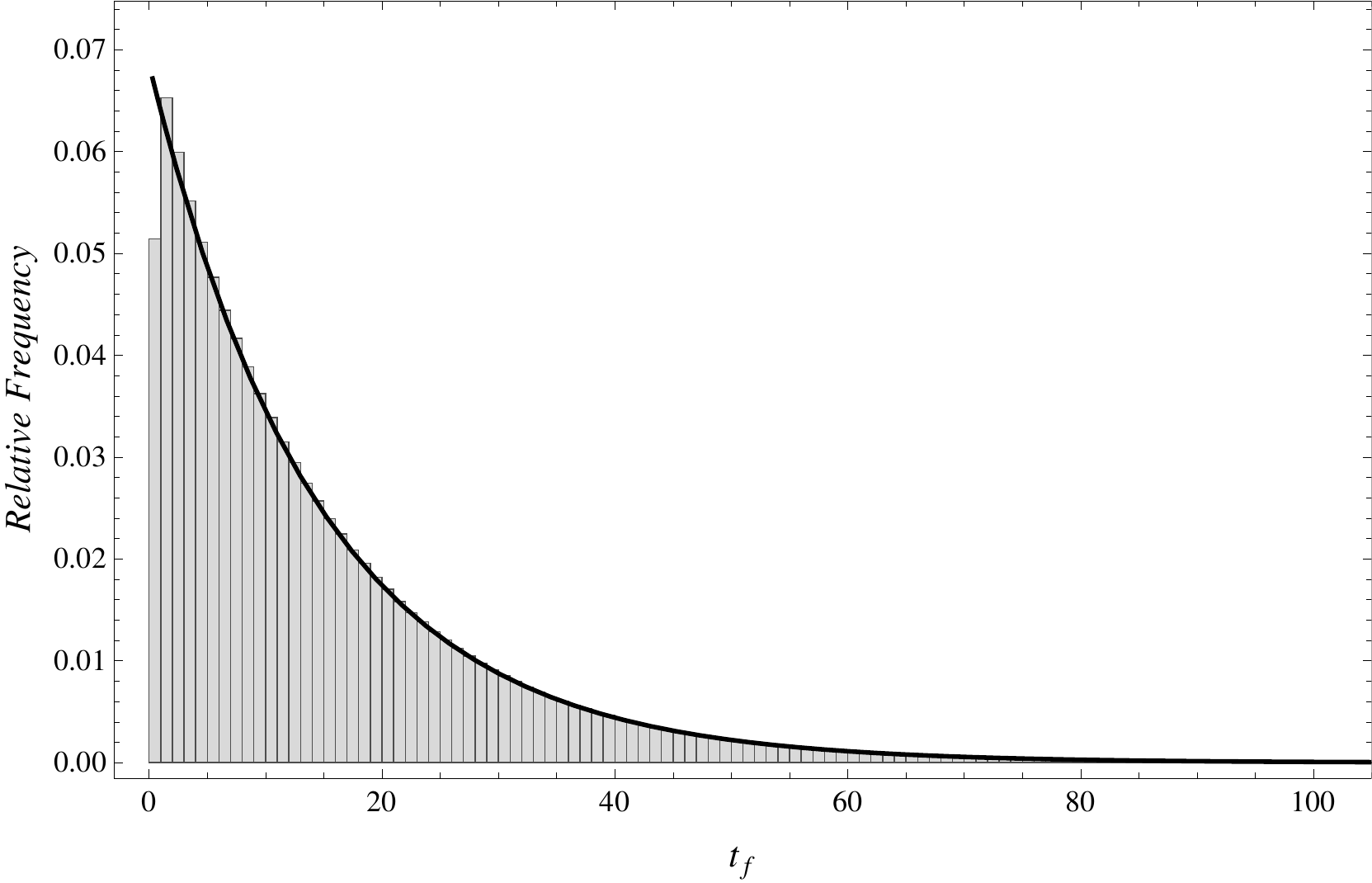} 
\caption{Probability of the expansion diverging to $-\infty$ as a
function of Planck time steps.  The solid line is the exponential
distribution (\ref{c1}).}
\label{f1}
\end{figure}

Using Mathematica \cite{Wolfram}, we have performed ten million 
runs of this simulation.  Figure \ref{f1} shows the probability density
of light cone ``collapse'' as a function of time in Planck units.  The 
mean time to collapse is $14.73\, t_P$, where $t_P$ is the Planck time; 
the standard deviation is $14.53\, t_P$.  These are perhaps large enough 
to justify our neglect of strong quantum gravitational effects, but 
small enough to probe the causal structure of spacetime in a 
physically very interesting region.

We see from the figure that the probability for ``collapse'' at the
first step is approximately $.051$.  This provides a useful check,
since this probability can be obtained directly from the distribution
(\ref{b5}): it is just the probability that the argument of the tangent 
in the collapsing branch of (\ref{b7}) is at least $\pi/2$ when 
$\lambda=1$, $\lambda_0=0$.  The exact result matches our 
simulation.

As shown in Figure \ref{f1}, after the first few steps the distribution 
can be fit very accurately to an exponential distribution:
\begin{align}
\mathop{Pr}(t_f=n) = \rho e^{-\rho n} \qquad
\hbox{with $\rho \approx .0686$} .
\label{c1}
\end{align}
This is surprisingly close to our approximation at the end of the
preceding section.  Moreover, while the details of individual runs
vary widely, almost all show $\bar\theta$ rising quickly to near its
maximum value of ${\bar\theta}_+$, confirming the starting 
assumption of our approximation.

Two-dimensional dilaton gravity thus appears to exhibit short
distance asymptotic silence, with vacuum fluctuations causing 
a rapid convergence of null cones.

\section{Generalizations and implications}

Our computations have been restricted to two-dimensional dilaton
gravity, primarily because this is the only setting in which the
probability distribution (\ref{b5}) is known.  We have also restricted 
ourselves to conformally invariant matter, for the same reason.
But the qualitative features of our results are largely independent
of such details, and we expect them to carry over to the full
four-dimensional theory.  In particular,
\begin{enumerate}
\item The vacuum fluctuations of the stress-energy tensor in four
dimensions are also expected to have a strict lower bound and an
infinitely long positive tail \cite{Fewster}.  If anything, the positive
tail seems to fall off more slowly than in two dimensions.  Moreover,
while our detailed results used a Gaussian test function to smear 
the stress-energy tensor, these features hold for any sufficiently 
compact test function \cite{Fewsterb}.
\item If the stress-energy tensor is bounded below, the expansion 
in four dimensions has bounds exactly analogous to those of (\ref{b7}),
even taking the same functional form \cite{Grant}.
\item Adding real matter would change the quantitative details of
our results.  But as long as that matter satisfies the null energy 
condition, it can only lead to further focusing.
\end{enumerate}

We thus expect something akin to short distance asymptotic silence
in four-dimensional general relativity.  We should, however, add three 
caveats.  First, we have treated our sequence of vacuum fluctuations  
as if they were statistically independent.  This is not quite 
right: the results of \cite{Fewster} imply that the (Gaussian smeared) 
stress-energy tensor $T_L$ at coordinate $u$ is weakly anticorrelated 
with the same object at $u+\tau$.  The correlation drops off sharply 
with distance, and should not affect our qualitative results.  It may 
be possible, though, to take this effect into account to produce a
more accurate quantitative picture.  Work on this question is in 
progress.

Second, the vacuum fluctuations found in \cite{Fewster} are 
fluctuations of the Minkowski vacuum (or, by conformal invariance, of 
any conformally equivalent vacuum).   While any curved spacetime is 
approximately flat at short enough distances, this is not enough to 
determine a unique vacuum, and we do not know how sensitive our results 
are to this choice.  

Third, we have by necessity neglected purely quantum gravitational 
effects, which could also compete with the vacuum fluctuations of 
matter.  This is not independent of the problem of choosing a vacuum; 
for instance, it is known that if one chooses the Unruh vacuum near 
a black hole horizon, the renormalized value of the shear term 
$\sigma_{ab}\sigma^{ab}$ in (\ref{a1}) is negative \cite{Candelas}.

If a proper handling of these caveats does not drastically change our 
conclusions, though, we have learned something very interesting about 
the small scale structure of spacetime.  The strong focusing of null
cones means that ``nearby'' neighborhoods of space are no longer in
causal contact.  This sort of breakup of the causal structure has been
studied in cosmology \cite{Uggla}, where it leads to BKL behavior
\cite {BKL}: each small neighborhood spends most of its time as an
anisotropically expanding Kasner space with essentially random 
expansion axes and speeds, but periodically undergoes a chaotic
``bounce'' to a new Kasner space with different axes and speeds.
It was argued in \cite{Carlip1,Carlip2} that such a local Kasner
behavior could explain the apparent spontaneous dimensional reduction 
of spacetime near the Planck scale, while preserving Lorentz invariance
at large scales.  Whether or not this proves to be the case, the short
distance collapse of light cones suggests both a new picture of 
spacetime and a new set of approximations for short distances.

\begin{acknowledgments}
S.\ Carlip  was supported in part by U.S.\ Department of Energy grant
DE-FG02-91ER40674.  J.~P.~M.\ Pitelli would like to thank Fapesp 
(grant 2008/01310-5) for financial support and the UC Davis
Physics Department for hospitality during this work.  R.~A.\ Mosna 
acknowledges support from CNPq and Fapesp.
\end{acknowledgments}

\end{document}